\documentclass[a4paper,11pt]{article}

\usepackage{jinstpub}

\usepackage[english]{babel}
\usepackage[utf8x]{inputenc}
\usepackage[T1]{fontenc}

\usepackage[doublespacing]{setspace}
\usepackage{amsmath}
\usepackage{graphicx}
\usepackage{subcaption}

\title{Investigation of the cryogenic scintillation of pure and doped sodium-iodine}

\author[a,1]{M. Clark,\note{Corresponding author.}}
\author[b]{F. Froborg,}
\author[a]{P.C.F. Di~Stefano}
\author[c]{and F. Calaprice}

\affiliation[a]{Department of Physics, Engineering Physics \& Astronomy, Queen’s University, Kingston, ON, K7L 3N6, Canada}
\affiliation[b]{Imperial College London, High Energy Physics, Blackett Laboratory, London SW7 2BZ, United Kingdom} 
\affiliation[c]{Physics Department, Princeton University, Princeton, NJ 08544, USA}

\emailAdd{12mtc2@queensu.ca}

\abstract{
We have studied the scintillation of pure and doped sodium iodide crystals handled in low humidity conditions under external $\alpha$ and $\gamma$ excitation from room temperature down to 4~K.
The light yield of pure sodium iodide was seen to increase at low temperatures by a factor 30 compared to room temperature, up to a maximum of 40~photons/keV under $\gamma$ excitation at 60~K, stabilizing to 30~photons/keV at lower temperatures.
Thallium doped sodium iodide fluctuates by 20\% around the room temperature value, stabilizing at 40~photons/keV at low temperature.
$\alpha / \gamma$ quenching factor stabilizes  at roughly 0.7 for both materials beneath 50~K. 
Time constants of both materials slow greatly at low temperature, reaching tens of microseconds.
Cryogenic applications of these materials are complexified by their mechanical fragility and hygroscopicity.}


\begin{document}

\maketitle
\flushbottom

\section{Introduction}

NaI(Tl) has been used as a scintillator for particle detection for many years~\cite{hofstadter_detection_1949}, and its wide-spread availability makes it a convenient low-cost choice for many particle-detection applications.
Undoped NaI has also been known to be an effective scintillator at low temperatures~\cite{van_sciver_fundamental_1958}.
The scintillation properties of these crystals have been characterized, though the difference between crystals can be large~\cite{sibczynski_properties_2011,moszynski_study_2003}, highlighting the importance of characterizing individual crystals in use, or at least batches of crystals.

The DAMA/LIBRA collaboration~\cite{bernabei_final_2013} detect an annual modulation signal~\cite{bernabei_final_2013,bernabei_first_2018} consistent with the presence of a dark matter halo, using an array of radiopure NaI(Tl) crystals.
To this date however, no other experiments have detected a similar signal with other target materials~\cite{kahlhoefer_model-independent_2018}.
The SABRE collaboration~\cite{antonello_sabre_2018,bignell_sabre_2020} aim to verify the DAMA/LIBRA annual modulation observation using the same target material in two distinct locations to remove seasonal effects: the Gran Sasso Lab in Italy and the Stawell Underground Physics Lab in Australia.

Though SABRE plans to use room-temperature NaI(Tl) crystals, others intend to use it a lower temperatures (for instance ASTAROTH~\cite{zani_astaroth_2021} at 87~K).
Moreover, advantages can be gained by introducing background discrimination in the experiment.
It has been shown that using an alkali halide scintillator at cryogenic temperatures, measuring both scintillation light and phonons created by particle interactions, can lead to the detection of a modulation in relatively small exposures~\cite{nadeau_sensitivity_2015}.
This allows the discrimination of electron recoils, usually from electron and $\gamma$ backgrounds, and nuclear recoils which are expected from WIMP interactions, reducing the level of background.
The COSINUS collaboration plans to use undoped NaI at cryogenic temperatures to search for dark matter signals~\cite{angloher_results_2017}.

This paper will present measurements made with SABRE-grown crystals to verify the scintillation properties of these NaI and NaI(Tl) crystals at low temperature.
It is important to ascertain these properties to allow for cryogenic experiments to properly design their light-collection and processing equipment.
It also may be interesting to determine the difference in performance for moderately cold (liquid nitrogen) temperatures to allow for better use for other applications.

\section{Experiment}

The experiment was carried out in a compact optical cryostat based on a Gifford-McMahon cryocooler capable of cooling small $5\times5\times2$~mm$^3$ samples of NaI and NaI(Tl) to any temperature between 3.5~K and 300~K~\cite{di_stefano_counting_2013,clark_particle_2018}.
The samples were taken from the tips of crystals grown by RMD Inc. as part of the SABRE collaboration~\cite{antonello_sabre_2018}.
These crystals were of a standard purity with an unknown crystal orientation and polished to transparency.
The NaI(Tl) sample is doped at roughly 1/1000 atoms Tl/NaI.

\begin{figure}
\centering
\includegraphics[width=0.6\textwidth]{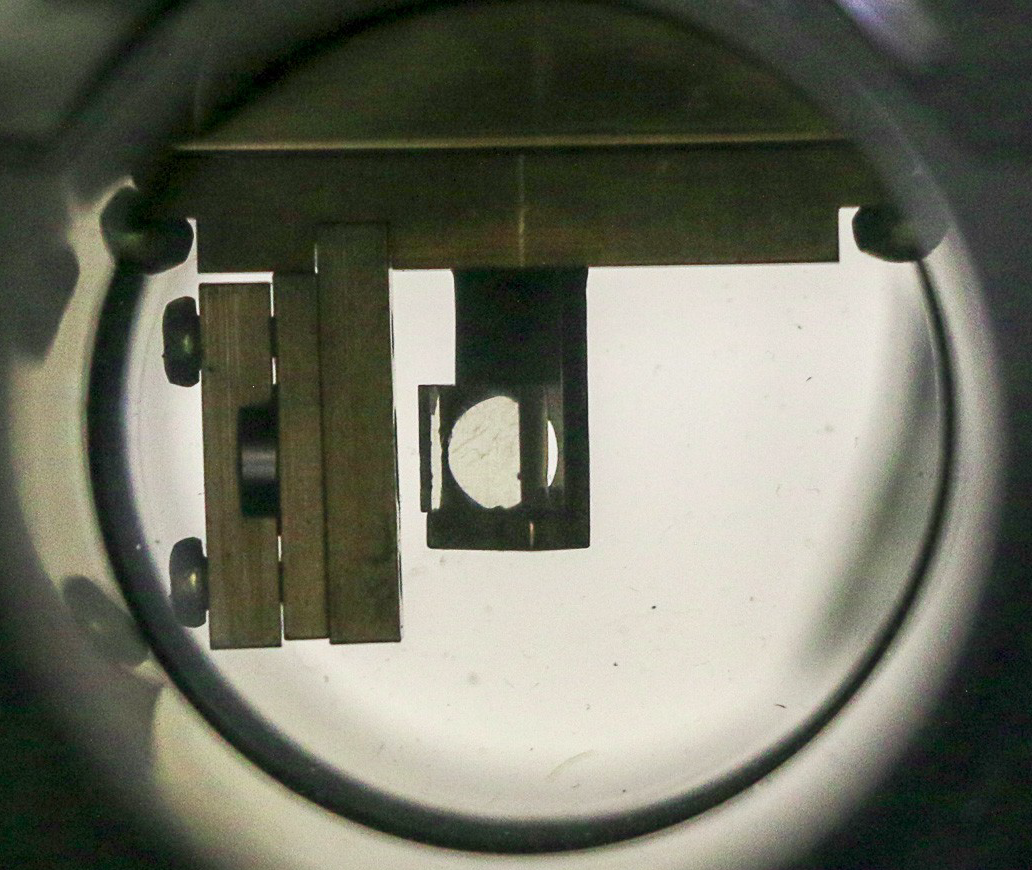}
\caption{\label{fig:cryst}Photograph of the NaI(Tl) sample mounted inside the cryostat. The crystal is held onto the gold-plated copper finger using silver paint for good thermal contact. A collimated Am-241 source illuminates the large face of the crystal.}
\end{figure}
Each crystal sample  was mounted inside the cryostat and excited by $\alpha$ particles and 60~keV $\gamma$ quanta from an internal collimated $^{241}$Am source mounted to the sample holder (Fig.~\ref{fig:cryst}).  
Additionally, an external $^{57}$Co source could be used to test $\gamma$ energies of 122~keV.
The $^{241}$Am $\alpha$ source used in this work was adapted from a common smoke detector where a protective film covers the radioactive material, reducing the energy of the $\alpha$-particles to 4.7~MeV.
The cryostat is housed in an acrylic glovebox to allow for careful control of the humidity.
By flowing pure nitrogen through the glovebox during installation of the crystal and throughout the data-taking, relative humidity within the glovebox was kept below 0.1\%, the sensitivity of our General Tools SAM990DW Psychrometer.

The samples were adhered to a custom-made holder using silver paint, instead of being mechanically held in place, to minimize mechanical stress as the sample cools. 
The crystal sample was at an angle of {30\textdegree} with respect to the $^{241}$Am source so that the emitted $\alpha$ particles were incident with one of the smooth, large faces of the sample, but at the same time this face was well exposed to a 28~mm diameter PMT (Hamamatsu R7056). A second PMT was placed opposite of the first one and was exposed to the rear face of the crystal, partially obstructed by the sample holder.
These PMTs have sensitivity to wavelengths down to 185~nm and a maximum quantum efficiency of 25\% at 420~nm. 
The PMT that was in view of the unobstructed face of the sample was used as the primary PMT for data acquisition, while the other was used to assist triggering. 
Based on the solid angle of the crystal exposed to the primary PMT and transmittance of the windows, we expect a light collection efficiency of roughly $10\%$ for the main PMT~\cite{nadeau_cryogenic_2015}. 
Note that no optical filters were used, so all light within the sensitivity of the PMT contributed to the light yield, as is standard in particle detection.

\begin{figure}
\centering
\includegraphics[width=\textwidth]{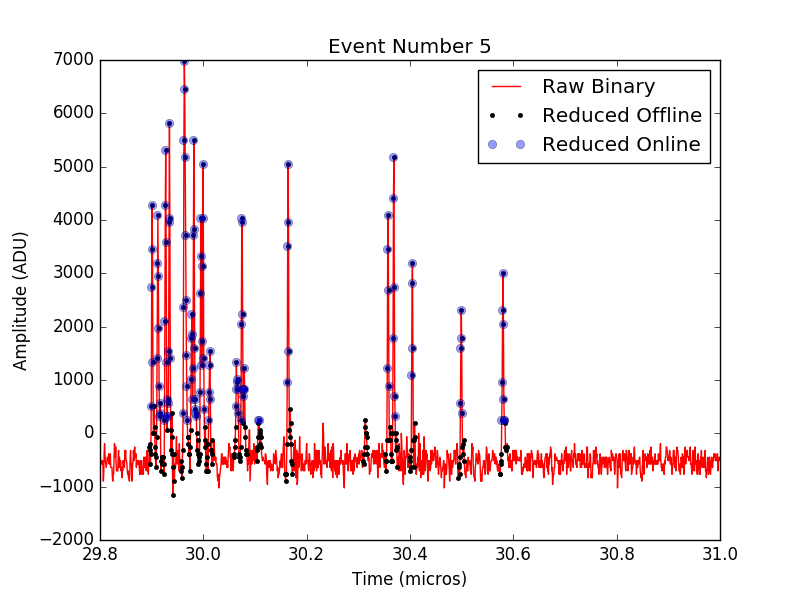}
\caption{\label{fig:pulse}An example of a single event in the raw digitizer trace (red line), recording only samples above the threshold (blue circles), and recording those plus the surrounding points (black dots).}
\end{figure}

The temperature of the crystal was controlled by a PID such that it was steady to within 0.1~K of the goal temperature during data taking.
Data were taken starting from room temperature, cooling down to the lowest temperature of 3.5~K in a single run to remove the possibility of releasing trapped energy through warming, known as thermoluminescence, which could occur when heating a crystal to a temperature that corresponds to a gap in its energy levels~\cite{sabharwal_thermoluminescence_1985}.
Scintillation pulse shapes were acquired using a modified version of the multiple photon counting coincidence technique~\cite{clark_particle_2018}. 
When a coincidence within 30~ns was detected between the two PMTs observing the sample, a 600~$\mu$s acquisition window sampled at $1.25 \times 10^9$~samples/second was opened in which the full digitizer trace was acquired.  The window included a 30~$\mu$s pre-trigger.
This trace was later zero-suppressed to reduce the data volume.
In previous measurements~\cite{clark_particle_2018}, we observed that simply recording information above a threshold caused a bias by not recording some information.
To combat this, we additionally record the 5 samples immediately preceding and following a sample above threshold.
This number was used since a typical single photon pulse was found to be around 10 samples wide. 
The threshold for a given pulse was set to 5 standard deviations of the fluctuations of the baseline of that pulse.
An example of the effect of this zero-suppression is shown in Figure~\ref{fig:pulse}.

The signal from the front PMT was passed to two digitizer channels with different ranges to allow for the measurement of both high voltage signals and small single-photon signals, as described previously~\cite{clark_particle_2018}.  
This dynamic range was important due to the large energy range of interest between 60~keV gammas and nearly 5~MeV alpha-particles.
The combination of the two channels resulted in a single measurement containing unsaturated, reconstructed pulses with higher detail at low voltages than would be possible with a single channel. 
At low temperatures where the light yield is higher (below 30~K for NaI(Tl) and below 150~K for NaI), a 10dB signal attenuator was used to prevent the data acquisition from saturating. 
The typical integral of a single photon was measured in both attenuated and non-attenuated configurations, such that the light yields can be directly compared in terms of number of photons. 
In addition, full datasets were taken at reference temperatures (50~K for NaI(Tl); 200~K and 150~K for NaI) in both configurations to ensure that the results were consistent.

\section{Data Analysis}

\begin{figure}
\centering
\begin{subfigure}{0.7\textwidth}
\includegraphics[width=\textwidth]{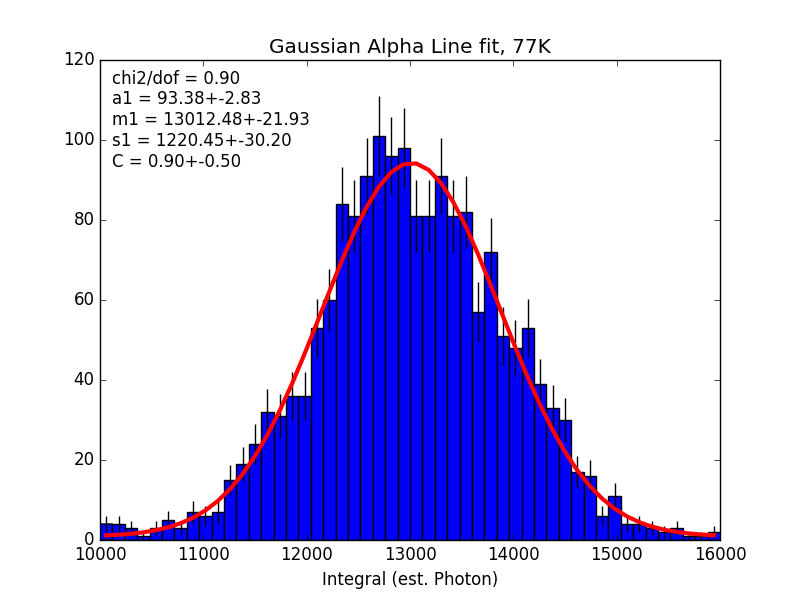}
\caption{\label{fig:alpha}4.7~MeV alpha events from Am-241}
\end{subfigure}
\begin{subfigure}{0.7\textwidth}
\includegraphics[width=\textwidth]{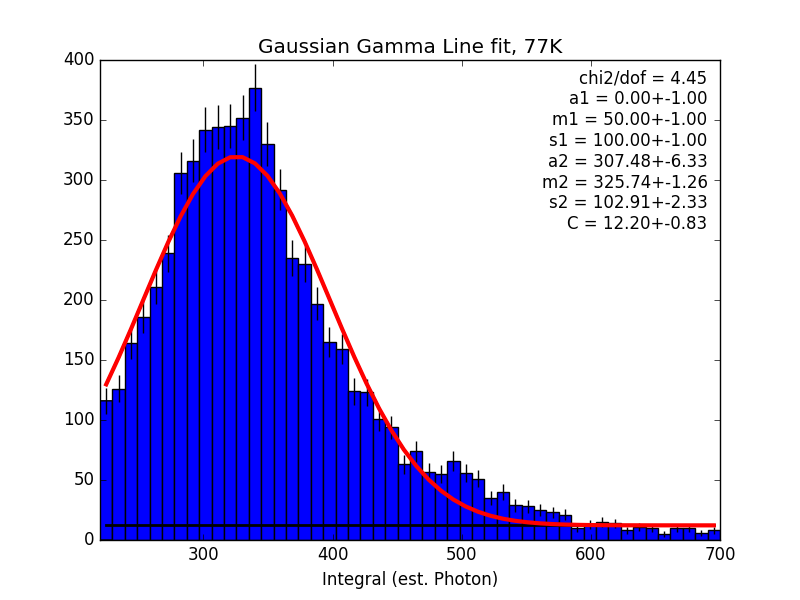}
\caption{\label{fig:gamma}60~keV gamma events from Am-241}
\end{subfigure}
\caption{\label{fig:peaks}Histograms of the measured light from events of both excitations from the Am-241 source for NaI at 77~K. In red is the fit to a gaussian curve, to determine the average light yield.  In the case of the gamma spectrum, the lower energy shoulder is attributed to Np X-rays from the source.}
\end{figure}

To determine the light yield, the integral of the events was binned and a gaussian fit was performed to find the peak corresponding to each particle excitation.
The mean of each peak was taken to be the average light yield of that excitation.
An example of an alpha excitation peak is shown in Figure~\ref{fig:alpha} and a gamma excitation peak in Figure~\ref{fig:gamma}.
The integral was then converted to a corresponding number of photo-electrons (PE) using the typical single photon integral in those digitizer conditions.
The conversion to number of PE allows us to compare our light yields with and without the signal attenuator directly, using different conversion factors for the two acquisition configurations.

It was noticed during the measurement of NaI that the alpha peak seemed to split into two distinct peaks at temperatures below 60~K.
It was later seen through inspection that the crystal was cracked after the experiment was completed, likely due to differences in thermal contraction between the copper sample holder and the NaI crystal.
We believe that the crack changed the optical properties of the crystal, changing the efficiency of detecting photons from events on either side of the crack.
Due to limited availability of crystal samples, this data will still be studied despite the change in crystal quality, with some added uncertainty after the cracking event.
To compare with the rest of our data, the weighted average of the two peaks was used to determine the light yield at temperatures where a split alpha peak is observed.

For each run, we perform a cut on the first photon time such that the first photon does not arrive before the expected coincidence window of 30~ns. 
This removes events that have photons in the pre-trigger, which could indicate that the tail of a previous event is present in our acquisition (pile-up event), which would give an inaccurate measurement of both the light yield and time behaviour. 
We then perform a cut on the light yield (integral of the event) to select for a particular particle interaction (4.7~MeV $\alpha$-particle, 60~keV $\gamma$). %
This is done using the gaussian fit as shown in Figure~\ref{fig:peaks} where all events within 2 sigma of the mean of the gaussian distribution are chosen to represent each population of events.
If the light yield is low, a minimum integral of two PE is set as the lower bound.
Finally, we perform a cut on the mean arrival time of photons after the first photon.  
A large mean arrival time compared to the typical value of the population could indicate another particle interaction occurred during the tail of the event which could interfere with our average pulses.

\section{Results and Discussion}

\subsection{Light yield}

\begin{figure}
\centering
\begin{subfigure}{0.7\textwidth}
\includegraphics[width=\textwidth]{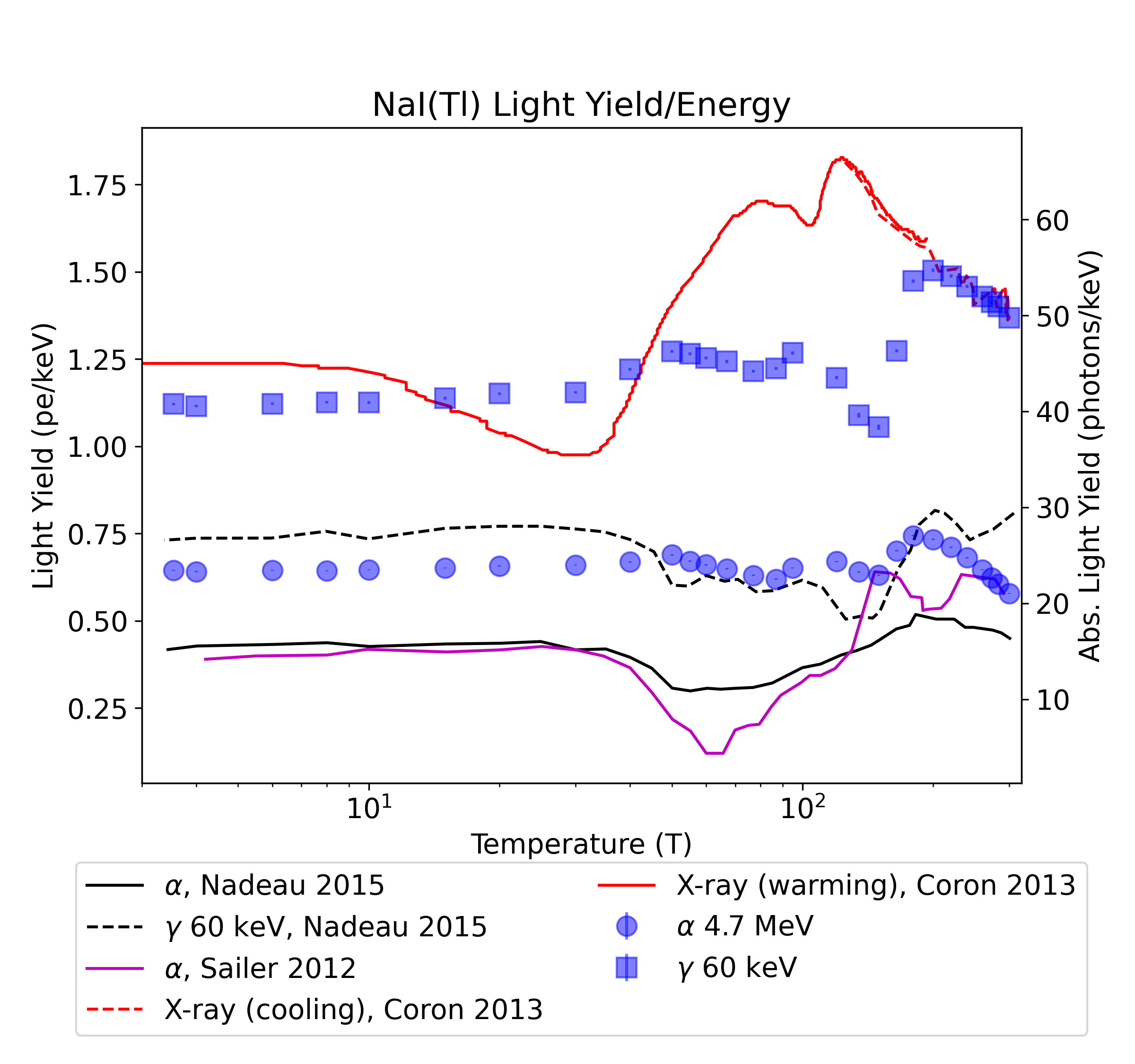}
\caption{\label{fig:NaITlLY}NaI(Tl)}
\end{subfigure}
\begin{subfigure}{0.7\textwidth}
\includegraphics[width=\textwidth]{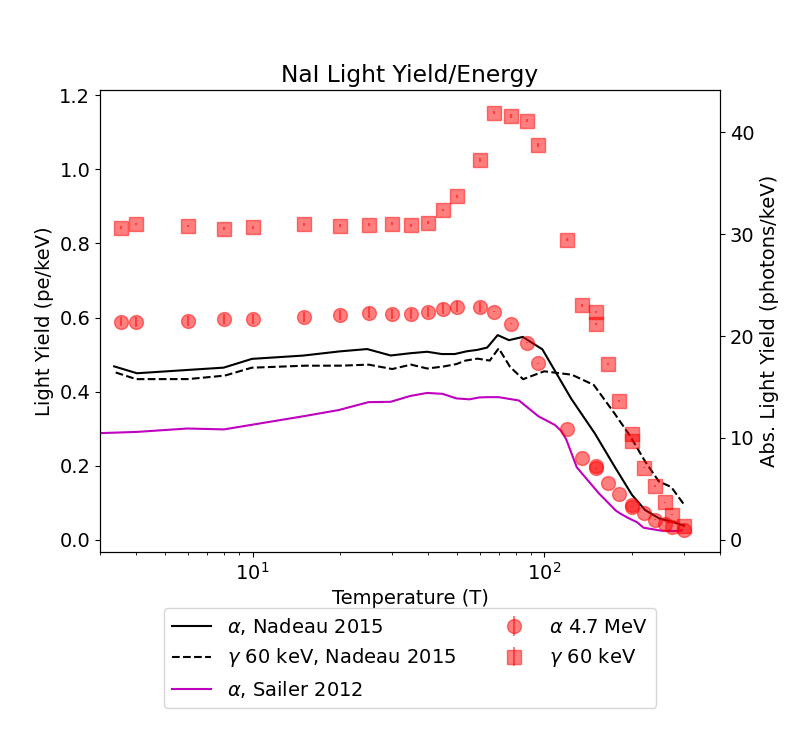}
\caption{\label{fig:NaILY}Undoped NaI}
\end{subfigure}
\caption{\label{fig:LYT}The light yield as a function of temperature for both crystals, along with comparisons to previous measurements~\cite{sailer_low_2012,nadeau_cryogenic_2015}, scaled to our measured value at room temperature. Data with signal attenuation are shown in empty markers, and without attenuation in full markers.  At temperatures where both attenuated and non-attenuated data was taken, results are consistent. Data from Coron (respectively Sailer) are normalized to our room temperature gamma (resp. alpha) value. Data from Coron 2013 are not time-resolved. Errorbars on each point are within the individual markers.} 
\end{figure}

The results of our light yield measurements as a function of temperature are shown in Figure~\ref{fig:LYT}, in addition to results from previous measurements~\cite{sailer_low_2012,nadeau_cryogenic_2015} that have been scaled to be equivalent with these results at room temperature for each type of excitation.
For NaI(Tl) (Fig~\ref{fig:NaITlLY}), the light yield changes by as much as 30\% as the temperature decreases for both $\alpha$ and $\gamma$ excitations.
At room temperature, we observe 1.4~PE/keV  (photoelectrons per keV) for the 60~keV $\gamma$-excitation, which, when accounting for the solid angle of our photomultiplier ($\approx$ 11\%) and the quantum efficiency (24\% at 400~nm), gives a rough estimate of $50 \pm 10$~photons/keV for the absolute light yield.
This is marginally higher than the frequently-quoted value of  38~photons/keV~\cite{holl_measurement_1988}.
At our lowest temperature of 3.5~K, we detect 1.1~PE/keV for $\gamma$ excitation corresponding to a decrease of light yield of 20\%.  This low-temperature value of $40 \pm 8$~photons/keV is slightly above the 29~photons/keV of CaWO4~\cite{mikhailik_performance_2010}.
The light yield of 4.7~MeV $\alpha$ excitations remain relatively stable at  0.65~pe/keV throughout the temperature range.
We observe a drop in light production at 150~K, which coincides with a thermoluminescence peak that has been observed previously for thallium-doped NaI~\cite{sabharwal_thermoluminescence_1985}.
This indicates there could be a shallow trap at the corresponding energy of $kT$ $\approx$ 10~meV.

In Figure~\ref{fig:NaITlLY}, we have compared the results reported in other studies~\cite{nadeau_cryogenic_2015,sailer_low_2012,coron_study_2013} to our own.
Since these results were reported as relative shifts in light yield, to compare with our results we have scaled each study to be equal to our own measurement at room temperature, for the specific particle interaction that was measured. We see features at approximately the same temperatures as other studies, though not to the same degree.
It should also be noted that Coron et al.~\cite{coron_study_2013} were not measuring time-resolved interactions but a constant illumination with X-rays, in addition to reporting in the warming phase over most of the temperature range, which could add additional complications from thermoluminescence.

In contrast, undoped NaI (Fig~\ref{fig:NaILY}) shows a large increase from  0.04~PE/keV at room temperature for $\gamma$ excitation, reaching a maximum of 1.15~PE/keV at 70~K, increasing by roughly a factor of 30.
The light yield then decreases, but remains high at 0.85~PE/keV until the lowest temperature, roughly $30 \pm 6$ photons/keV in absolute light yield. 
This is a much higher relative increase than has been seen previously, with previous measurements increasing by at most a factor of 10~\cite{nadeau_cryogenic_2015}.

It should be noted that the decrease in observed light yield coincides with the temperature at which the crystal is suspected to have cracked.
This may indicate that thermal stresses within the crystal, which were released when the crystal cracked, had an effect on the light yield. Another factor may be changes in optical efficiency.

\subsection{Quenching factor}

\begin{figure}
\centering
\includegraphics[width=0.7\textwidth]{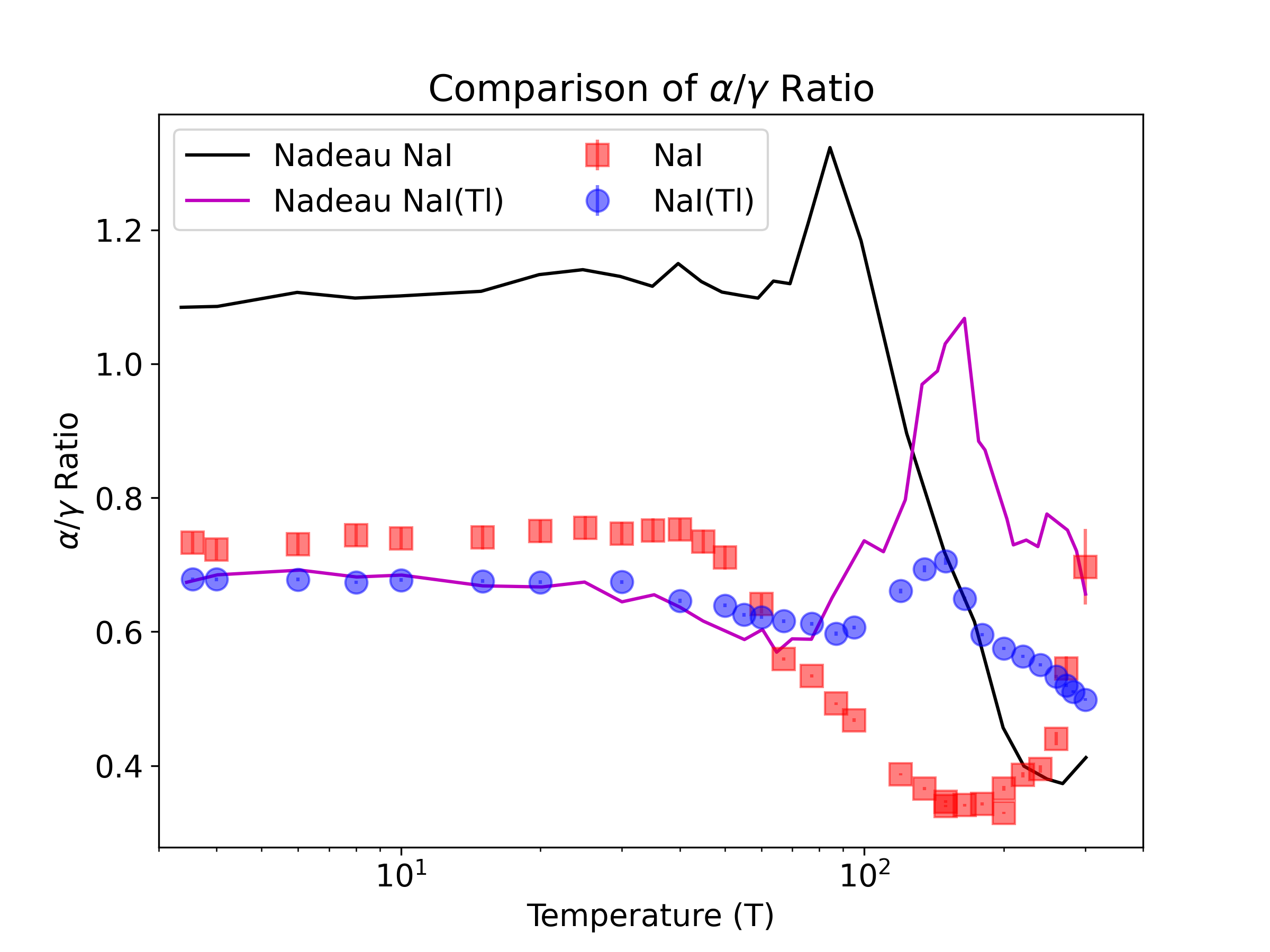}
\caption{\label{fig:ag}Measured $\alpha$/$\gamma$ ratio for both NaI and NaI(Tl) as a function of temperature.  Errorbars on each point are within the individual markers.}
\end{figure}

From the previous measurements, we have also determined the $\alpha$/$\gamma$ ratio of both NaI and NaI(Tl) as a function of temperature.
The results are shown in Figure~\ref{fig:ag}.
Since these results compare $\alpha$ and $\gamma$ excitations of different energies, we have employed a theory of inorganic scintillator non-proportionality to correct for this, transforming the 60~keV $\gamma$ light yield to that of a 4.7~MeV $\gamma$, as has been described in previous publications~\cite{clark_particle_2018}.

NaI(Tl) shows a slight increase with decreasing temperature, with a value of 0.7 at 3.5~K.
A peak value of 0.7 is also seen at the same location in the dip in light yield, corresponding again with the likely location of a shallow trap at the corresponding energy~\cite{sabharwal_thermoluminescence_1985}.

Undoped NaI also shows an evolution over temperature, but in this case drops from 0.7 at room temperature to 0.35 at 160~K, returning to 0.7 at 40~K once again.
This result differs from our previous measurements~\cite{nadeau_cryogenic_2015}, in  which the relative humidity was only kept below 20\% during sample handling, as opposed to the more stringent upper limit of 0.1\% in the present work.  This indicates that light yield of surface events, caused by alphas, depends on the state of the sample surface.  An alternative explanation is that the discrepancy is caused by differences in trace contamination between the different lots of samples.

\subsection{Time Structure}

\begin{figure}
\centering
\includegraphics[width=1\textwidth]{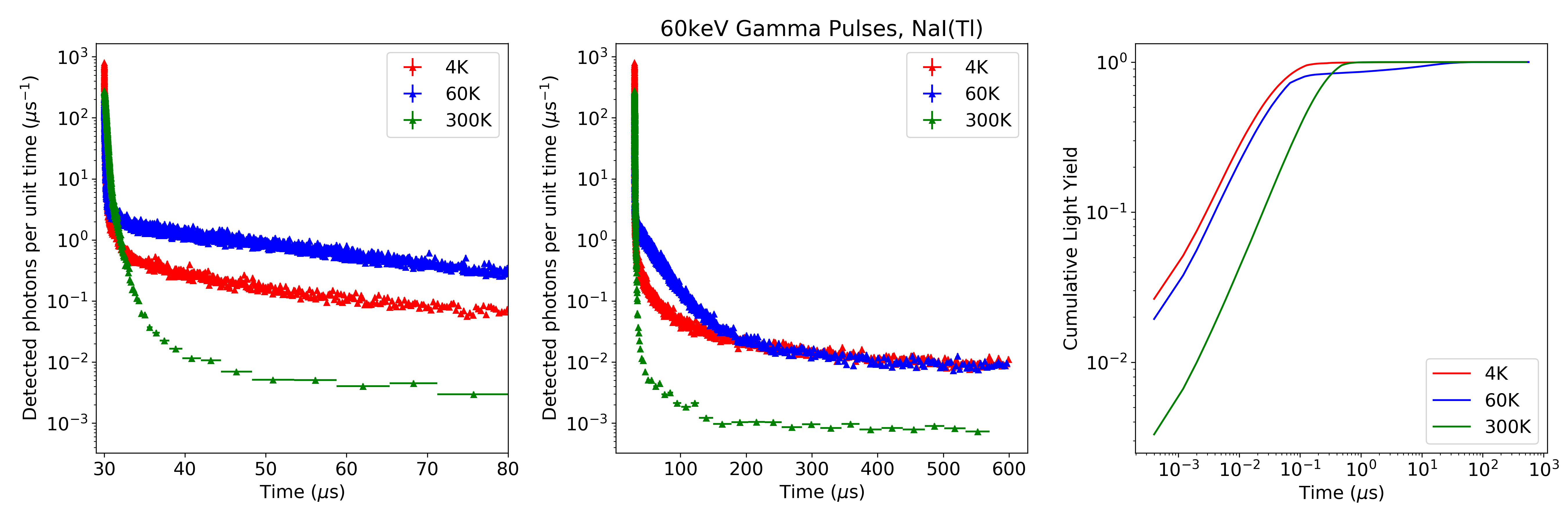} \\
\includegraphics[width=1\textwidth]{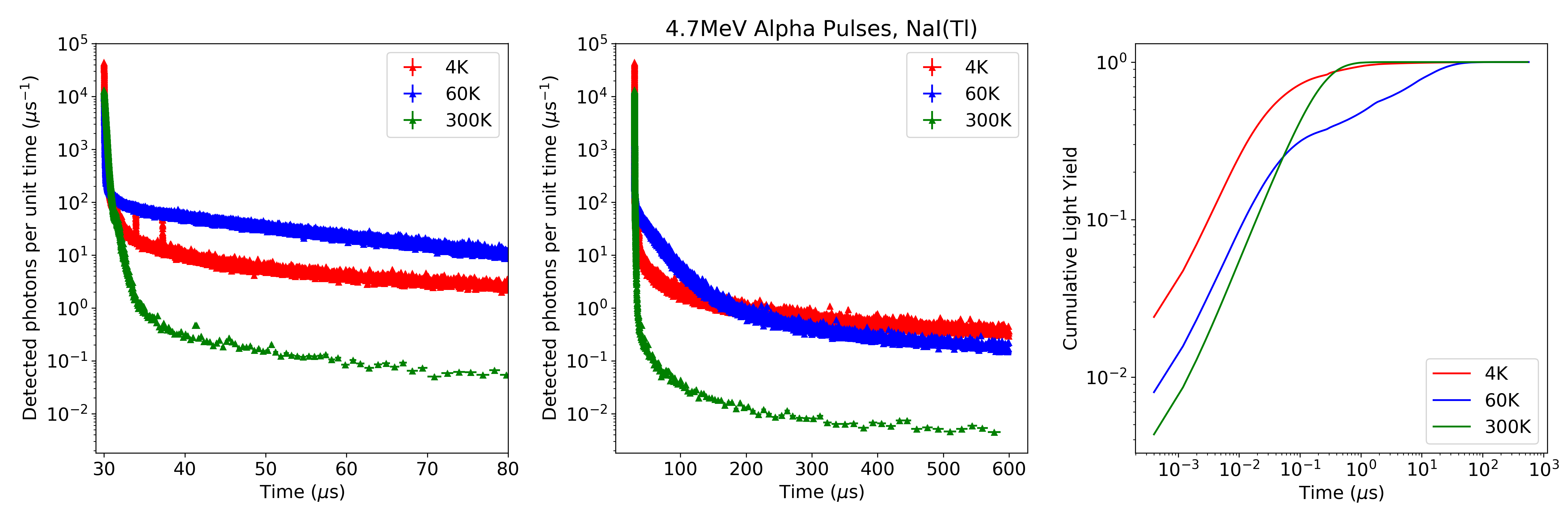} \\
\includegraphics[width=1\textwidth]{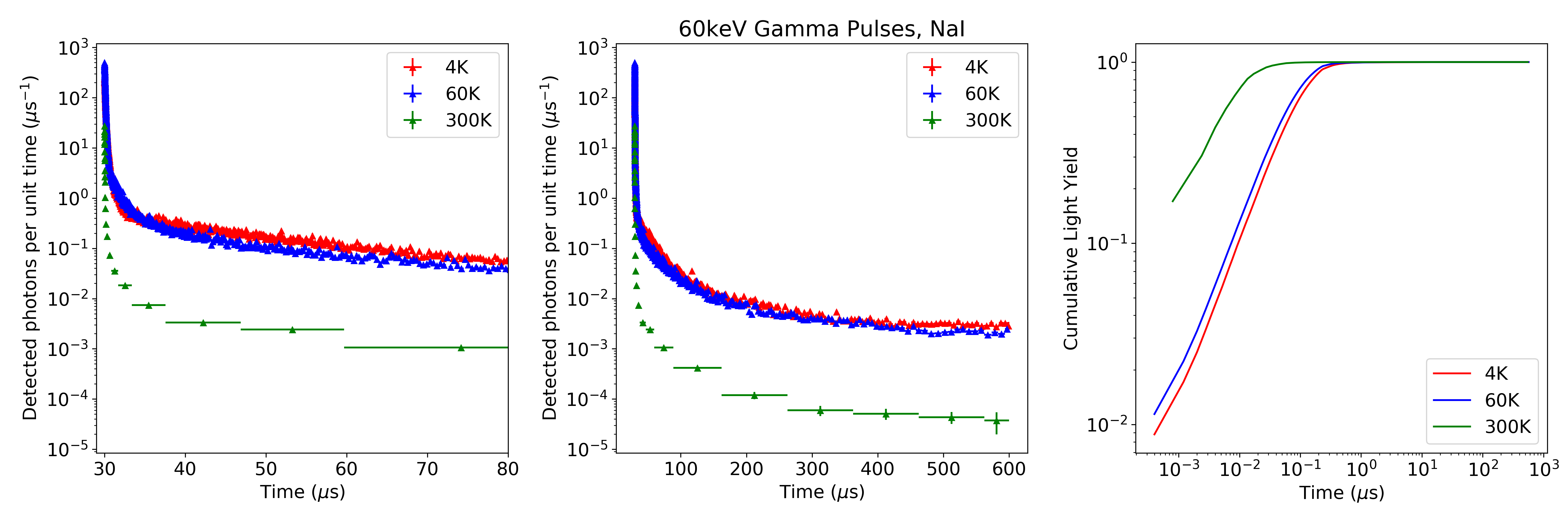} \\
\includegraphics[width=1\textwidth]{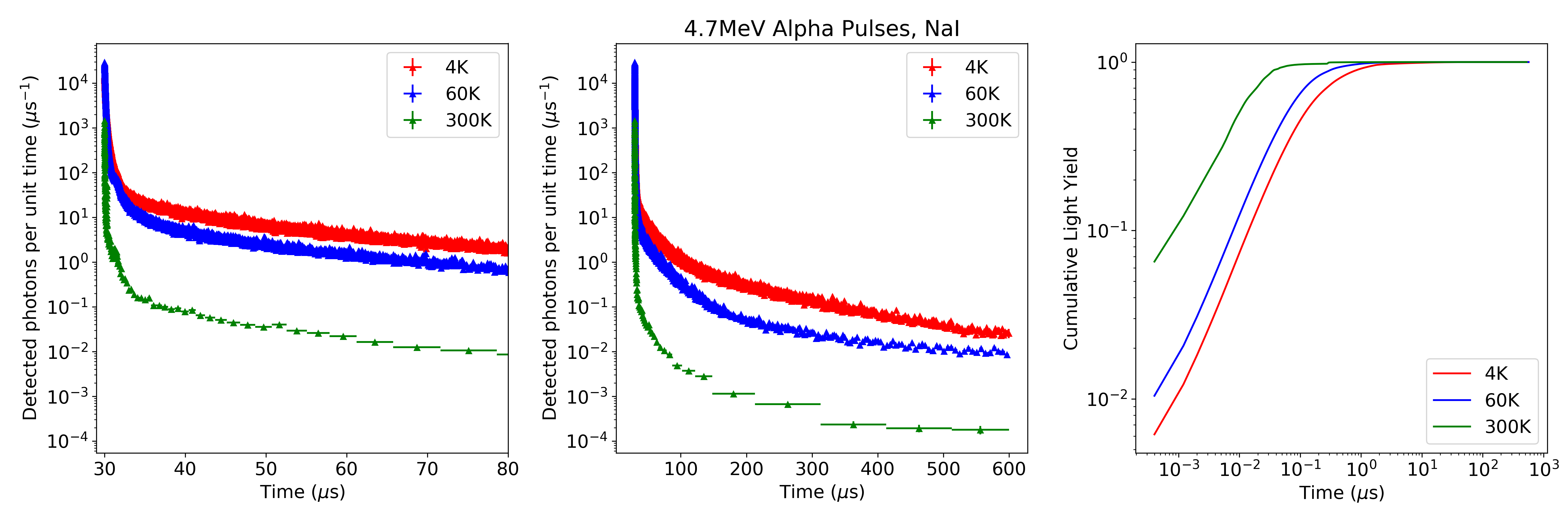}
\caption{\label{fig:AllPulses_cum}The average pulses from NaI(Tl) (top two rows) and NaI (bottom two rows) (Run 83) under $\gamma$ excitation (odd rows)  and $\alpha$ excitation (even rows), at 3 different temperatures. Pretrigger is 30~$\mu$s long. Trend is for pulses to become longer as temperature decreases. }
\end{figure}

\begin{figure}
\centering
\includegraphics[width=0.8\textwidth]{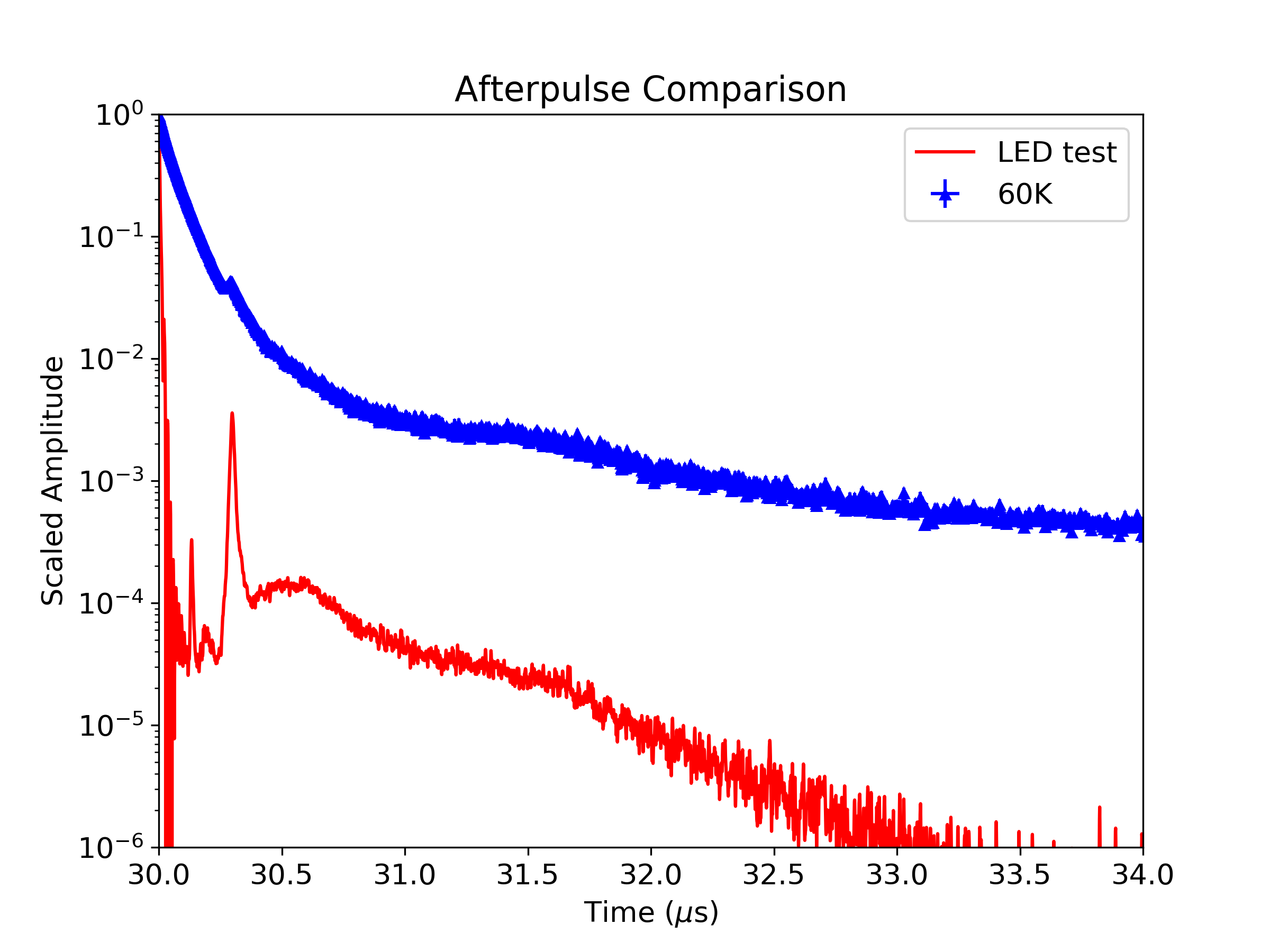}
\caption{\label{fig:aftpcomp}The LED measurement of afterpulsing overlaid with data from $\alpha$ excitation in NaI at 60~K. The largest afterpulse corresponds with the pulse shape at low time, and a later bump may correspond with the later bump in the pulse shape.  Pulses have been shifted to correspond to a pretrigger of 30~$\mu$s.}
\end{figure}

We have also investigated the pulse shapes induced by alpha and gamma interactions at different temperatures (Fig.~\ref{fig:AllPulses_cum}).  The time constants tend to increase as the temperature decreases, as observed previously in various materials (eg~\cite{verdier2011scintillation}) including NaI~\cite{birks_theory_1964}.  
A more precise determination of the time constants is complicated by the presence of afterpulsing in the PMs, which has been independently confirmed by exposing the PMs to brief light pulses from an LED (Fig.~\ref{fig:aftpcomp}).  At the standard operating voltage of the PMs, a main narrow structure was observed $\sim 300$~ns after the excitation, as was a broad secondary feature $\sim 1.5$~$\mu$s after the excitation.  This corresponds with the features in the scintillation pulse, as can be seen in Fig.~\ref{fig:aftpcomp}.
The integral of the afterpulses represents 3\% of the signal from the LED, and thus has a small effect on the light yield measurement.  
Attempts to deconvolve the afterpulsing from the overall pulse were hindered by the fact that the afterpulsing appears to have increased in the 2 years between the main experiment and the tests.

\begin{figure}
\centering
\includegraphics[width=0.45\textwidth]{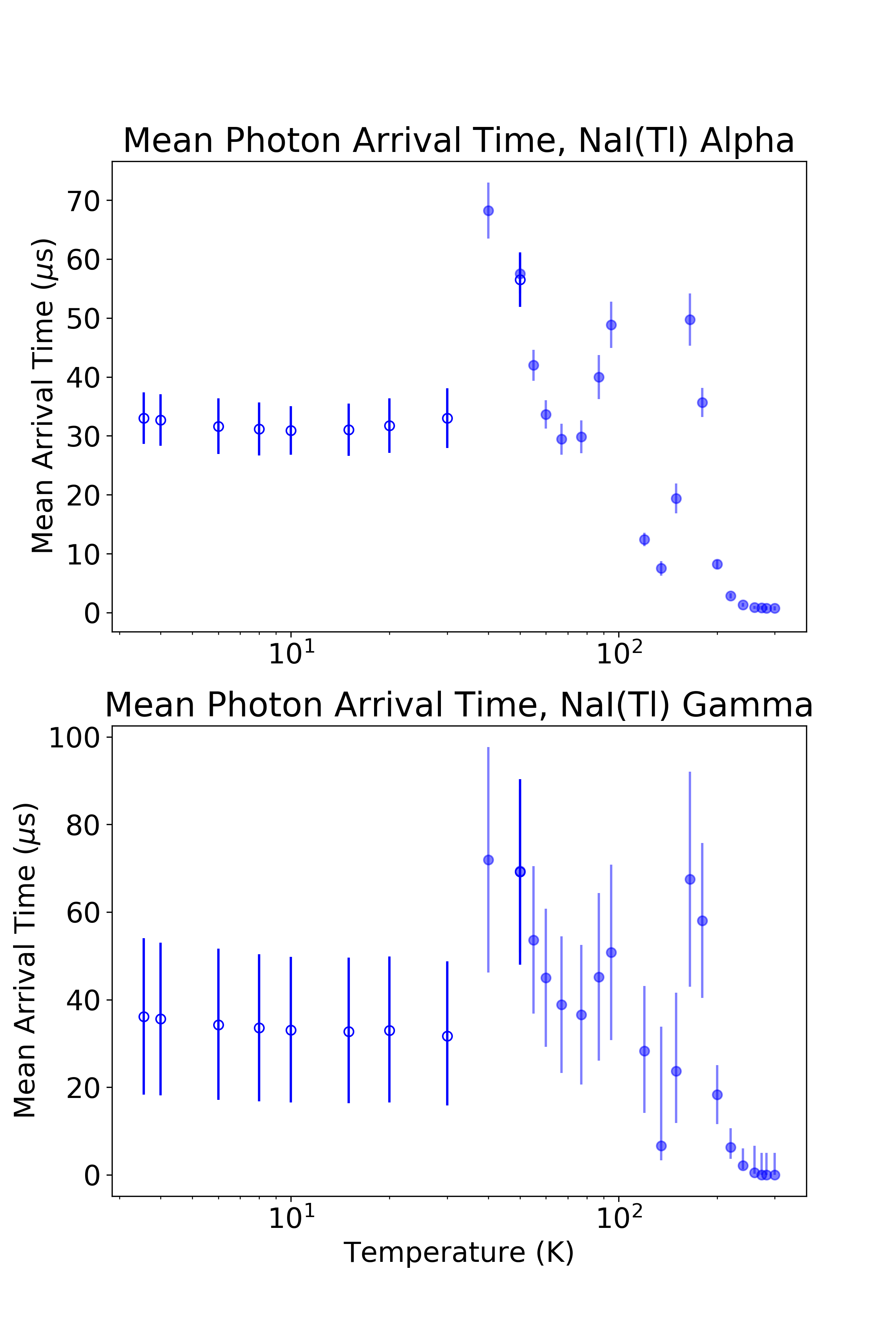}
\includegraphics[width=0.45\textwidth]{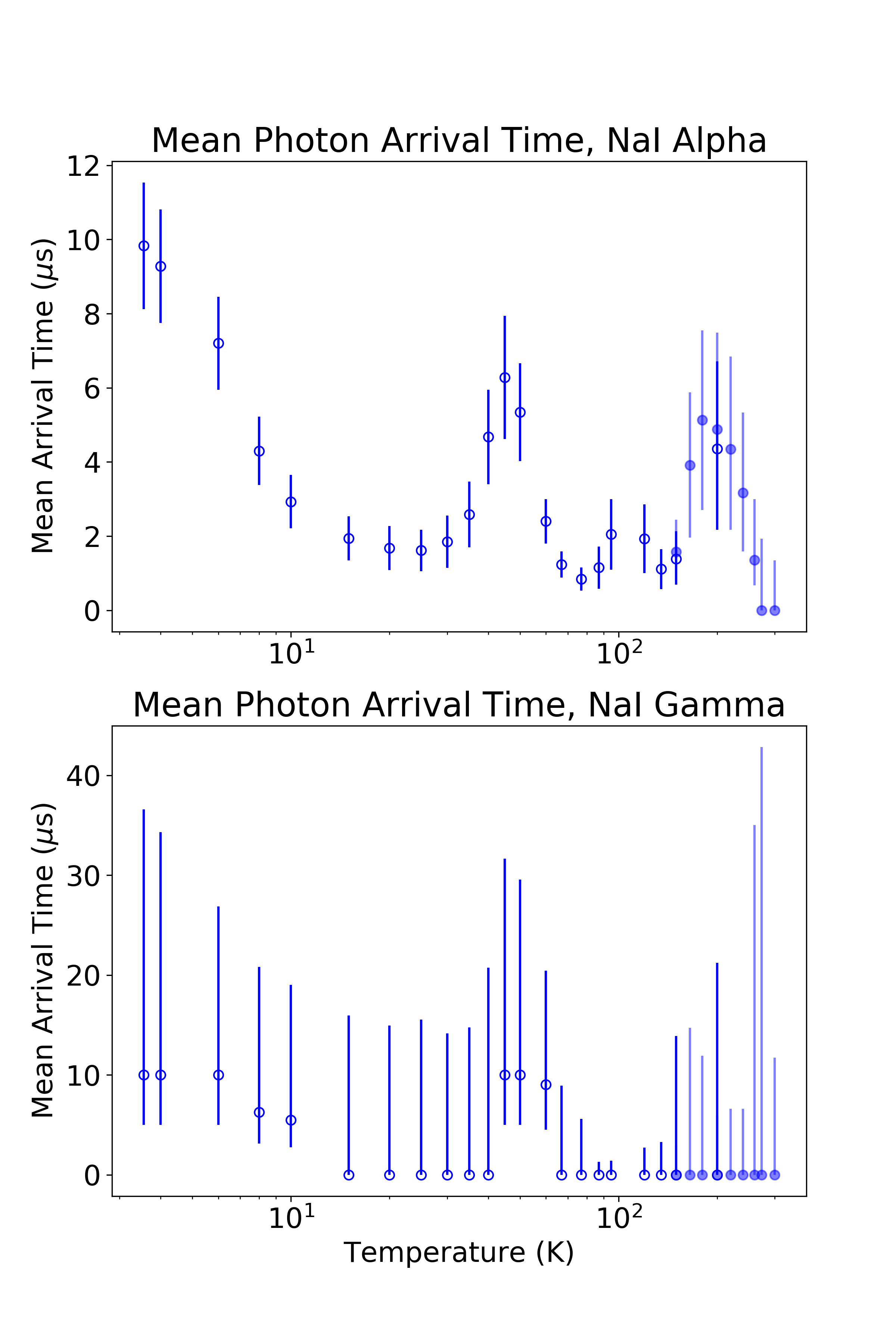}
\caption{\label{fig:allMAT}The evolution of mean photon arrival time as a function of temperature for the NaI(Tl) sample (left) and NaI sample (right).  Filled (respectively empty) markers are data taken without (with) attenuator. For NaI(Tl), cross check at 50~K showed consistent results with and without attenuator.  For NaI, cross checks, at 150~K and 200~K, were also consistent. }
\end{figure}

To present a quantitative measure of the changing time structure as a function of temperature, Figure~\ref{fig:allMAT} shows the mean arrival time of photons after the first detected photon, for $\alpha$ and $\gamma$ events for each crystal.  
For each event of a given type of particle (alpha or gamma), we calculate the mean arrival time of photons after the first one; the standard deviation of these values for a given population are used as error bars.
Unusually, the slowing trend in the data as the temperature is cooled is not monotonic.  All the data seem to show particularly long pulses at $\sim 40$~K and $\sim 175$~K.

We considered if the attenuator, used at low temperatures, could have played a role in this, since it might have made it harder to detect straggling single photons that appear at the low temperature.  Checks were therefore carried out with and without attenuator at 50~K for NaI(Tl) and at 150~K and 200~K for NaI; results were found to be consistent, making the attenuator an unlikely explanation.
We also note that the attenuator would have had little effect on the light yield, because for both samples at all temperatures, the bulk of the photons are not resolved individually.

\section{Conclusion}

In this work, we have shown that the light yield of the NaI and NaI(Tl) crystals grown by the SABRE collaboration perform very well in terms of scintillation yield at low temperatures.
NaI(Tl) maintains its high light yield at all measured temperatures, while undoped NaI sees a large increase in light yield as temperature decreases, up to a factor of 30 times higher than at room temperature.
The evolution of the light yield of NaI(Tl) as a function of temperature is quite irregular above $\sim 50$~K, which is consistent with previous experiments.
For gamma excitations at 4~K, compared to CaWO4, the light yield of NaI(Tl) is slightly greater while that of NaI is comparable.
Because of our instrumentation and relatively high light collection, we are able to measure both $\alpha$ and $\gamma$ excitations under the same conditions.
We have reported here the $\alpha$/$\gamma$ ratio as a function of temperature, showing that it changes over the temperature range, with a maximum value of 0.7.  
The response to alphas may very well depend on the state of the surface of the crystal, which itself depends on how much humidity the samples have been exposed to. 
Therefore, care was taken to handle these samples under dry nitrogen to maintain a good surface quality.
In addition, all measurements of a given sample were made during a single cooling cycle to avoid contributions from thermoluminescence.
Another factor that may be relevant is the presence of the impurities in a given sample.
These various factors could explain the differences between this result and previous measurements in light yield and $\alpha$/$\gamma$ ratio.

Time-resolved measurements of the scintillation light show presence of  light out to 600~$\mu$s after the initial excitation. At cryogenic temperatures, undoped NaI shows an increased proportion of light at long times, where NaI(Tl) shows a larger increase at short times, even if the absolute light yield at long times increases.  The mean arrival times of photons from both $\alpha$ and $\gamma$ excitations do not increase monotonically, showing a complicated evolution as the temperature decreases.

Significant efforts were made to shelter the samples from humidity and to handle them delicately to reduce thermal stress during cooling. The sensitivity of NaI, both pure and doped, to humidity and the mechanical fragility of the crystals underscore the challenges of using these materials in cryogenic applications, though the potential for background discrimination may make it worth the effort.

\acknowledgments
Funding in Canada has been provided by NSERC through SAPIN grants, by CFI-LOF and ORF-SIF.
Queen's summer student Miaofen Zheng contributed to the data taking.  Queen's Visiting Research Student Ilian Moundib contributed to the afterpulsing analysis.
F.~Froborg has received funding from the European Unions Horizon 2020 research and innovation programme under the Marie Sklodowska-Curie grant agreement No 703650.  F.~Froborg's participation in the experiment was made possible by a grant from the Queen's Principal's Development Fund.
We thank Radiation Monitoring Devices (RMD) and the SABRE Collaboration for providing the test samples.

\bibliographystyle{JHEP}
\bibliography{NaIPaper}

\end{document}